# Noncommutative Minisuperspace Cosmology

Aleksandar Zejak[*)]
*Mathematical Institute SANU, Belgrade, Serbia*

Branko Dragovich
*Institute of Physics, Belgrade, Serbia*

**Abstract**: We consider the noncommutative minisuperspace classical and quantum cosmologies.

**Introduction**

The considerations of noncommutative minisuperspace cosmologies have recently attracted much attention [1-5] due to the fact that the landscape of string theory vacua [6], i.e. the number of string theory vacua with positive energy densities, is too big [7-11], and that within the string theory itself there is no reasonable selection mechanism between them. Also, in some studies [4], [5], a solution of the cosmological constant problem was suggested. Interest in canonical noncommutativity and its application to minisuperspace models was stimulated by the discovery of its relation to the string and M-theory [12-15]. This kind of noncommutativity arises in the low energy effective field theory on a D-brane in a constant antisymmetric background B-field. Also, there are many proposals for a noncommutative theory of gravity [16-37], in which in defining the noncommutative minisuperspace models the noncommutative parameter corresponding to space-time noncommutativity may be treated in a new way, though the cosmology of these models depends on time parameter only [1]. The noncommutativity parameter $\theta$, which we shall introduce in this work, may then be considered as the effective noncommutativity parameter, and the noncommutative minisuperspace models may therefore serve as toy models which test the selective role of this effective noncommutativity in the choice of the vacuum, or of the wave function as the ground state of the universe in such a quantum cosmology. It is well known that the Hartle-Hawking condition [38], [39] does not fix the wave function of the universe uniquely [40], [41], what additionally justifies the consideration of the influence of noncommutativity on this condition via the noncommutative minisuperspace quantum cosmology.

---

[*)] e-mail: acazejak@yahoo.com

In this work we consider the canonical noncommutativity of the 2D minisuperspace model of the homogeneous and isotropic closed universe with the homogeneous dimensionless scalar matter field $\phi$, with the potential of the form $\alpha \operatorname{ch} 2\phi + \beta \operatorname{sh} 2\phi$, where $\alpha$ and $\beta$ are arbitrary real parameters. Values of these parameters determine the class of the potentials as well as of the corresponding universes. The standard version of this model has already been considered by Halliwell et.al. [42] who concluded that the suggested potential is not void of physical meaning. In this (standard) case, for $\alpha = \beta$, the matter potential yields the power-law inflation model, while for $\alpha > 0$ and $\beta = 0$, and $\phi = 0$, it yields the standard de Sitter cosmology. We shall apply the canonical noncommutativity of the linear minisuperspace coordinates $x = a^2 \operatorname{ch} 2\phi$ and $y = a^2 \operatorname{sh} 2\phi$ (where $a$ is the dimensionless cosmic scale factor), where we shall, motivated by physical arguments, have to seriously consider the noncommutativity of curvilinear minisuperspace coordinates $a$ and $\phi$. As the main problem in the development of the theory of gravitation on noncommutative manifolds is the implementation of general coordinate covariance (finite diffeomorphisms), the definition of derivatives and the metricity condition [43], [44], we shall in the case of the noncommutative minisuperspace, as an alternative to other possibilities [45-48], try to overcome this problem by modifying the Moyal product [49]. The price that we shall have to pay in doing this is the change of the algebraic structure of the noncommutative minisuperspace at the expense of its "relative" noncommutative geometry. Also, the procedure will be adapted to the Weyl rescaling, which appears in ours as well as in the standard (commutative) model [42], [50]. This will enable the relation of the Moyal product to its modification as introduced here, and will justify the introduction of this formalism via the classical and quantum noncommutative Hamiltonian dynamics of our model. By using the Moyal approach it will be shown that at the quantum level the noncommutativity results in the appearance of the $\theta$-phase factor and the $\theta^2$ containing term in the argument of both the general and particular solutions of the corresponding noncommutative Wheeler-DeWitt equation. It will turn out that the $\theta$-phase factor plays significant role in interference phenomena, while the $\theta^2$ term in the semiclassical approximation for extends the classically forbidden region of the minisuperspace. It also appears that at the Hartle-Hawking condition and for different choices of the gauge condition $\dot{N} = 0$ ($N$ is the lapse function) this $\theta^2$ term either decreases or increases the semiclassical probability amplitude for tunneling from nothing to the closed universe with the stable matter potential. This analysis already demonstrates the significant superselection role of noncommutativity not only in the choice of the corresponding gauge condition but also in the choice of the most probable semiclassical ground state of the universe. Finally, it will be shown that under the Hartle-Hawking condition and when $\alpha > 0$ and $\beta = 0$ the canonical noncommutativity of the minisuperspace prefers as the most probable the creation of the closed universe with $\phi = 0$ by the semiclassical wave function which for $\theta = 0$ corresponds to the geometry of filling in the three-sphere with more than half of a four-sphere of radius $\sqrt{(1/\alpha)}$ [38], [39]. It also appears that for the same case at the classical level this kind of noncommutativity of the minisuperspace explicitly determines the cosmic time dependence of the cosmological "constant" $\Lambda_{eff}(t)$ the value of which is smaller than the standard cosmological constant at all times, except in the infinite future.

**Noncommutative model of the closed Universe with the scalar field**

We consider the model of the closed universe that is described by the rescaled Friedmann-Robertson-Walker (FRW) metric determined by the space-time interval in the form [42]

$$ds^2 = \sigma^2 \left[ -\frac{N^2(t)}{a^2(t)} dt^2 + a^2(t) d\Omega_3^2 \right], \qquad (1)$$

where $N(t)$ is a dimensionless lapse function, $a(t)$ is the dimensionless cosmic scale factor, $\sigma^2 = \kappa^2/3\mathbf{V}$ ($\kappa^2 = 4\pi G$, G-Newton's constant, c = 1 (c-speed of light) and $\mathbf{V} = 2\pi^2$ is the volume of the unit 3-sphere) and $d\Omega_3^2 = d\chi^2 + \sin^2\chi (d\vartheta^2 + \sin^2\vartheta\, d\varphi^2)$ is the metric of the unit 3-sphere. As the cosmological models with the Coleman-Weinberg potential can not be exactly solved we make use of the alternative redefined potential of the scalar field $U(\phi)$ in the form:

$$U(\phi) = \alpha\, \text{ch}\, 2\phi + \beta\, \text{sh}\, 2\phi \equiv \frac{2\kappa^2\sigma^2}{3} V(\Phi), \qquad (2)$$

where $\alpha$ and $\beta$ are arbitrary real parameters, and the dimensionless field $\phi$ is related to the ordinary scalar field $\Phi$ as follows:

$$\Phi = \phi\sqrt{3}/\kappa = \phi/\pi\sigma\sqrt{2}. \qquad (3)$$

The corresponding minisuperspace model is now 2D and is described by the cosmic scale factor $a$ and scalar field $\phi$. From (1) and (2) there follows the form of the metric of a 2D minisuperspace model of the closed universe and rescaled potential [51]:

$$G_{AB}(q) dq^A dq^B = \sigma(-a^2 da^2 + a^4 d\phi^2),$$
$$\mathbf{U}(q) = \frac{1}{2\sigma}[a^2 U(\phi) - 1], \qquad (4)$$

where $q^A \equiv (a, \phi)$. Halliwell [50] demonstrated that for any nD minisuperspace model the ordering parameter $\xi$ in the conformally invariant Wheeler-DeWitt (WDW) equation (Planck's constant $\hbar = 1$):

$$H\Psi(q^A) = \left[ -\frac{1}{2}\Delta + \xi R + \mathbf{U}(q) \right] \Psi(q^A) = 0 \qquad (5)$$

(where the Laplace-Beltrami operator is $\Delta \equiv \frac{1}{\sqrt{-G(q)}} \frac{\partial}{\partial q^A}\left[\sqrt{-G(q)}\, G^{AB}(q) \frac{\partial}{\partial q^B}\right]$, $R$ is the scalar Riemannian curvature in minisuperspace $M$, $U(q)$ is the minisuperspace potential and $\Psi(q^A)$ is the wave function of the universe) equals $\xi = (2-n)/[8(1-n)]$ for $n \geq 2$. In our conformal model of the universe this parameter equals zero. Also, from metric (4) there follows that Riemannian curvature tensor of the minisuperspace $M$ is zero, what in turn implies that the application of the smooth transformation:

$$q^A \equiv (a, \phi) \rightarrow \tilde{q}^A \equiv (x, y) \qquad (6)$$

with

$$x = a^2 \, \text{ch}\, 2\phi, \quad y = a^2 \, \text{sh}\, 2\phi \qquad (7)$$

yields the metric of the flat minisuperspace in the form:

$$\tilde{G}_{AB} d\tilde{q}^A d\tilde{q}^B = \frac{\sigma}{4}(-dx^2 + dy^2) \qquad (8)$$

and the minisuperspace potential in the form:

$$\tilde{U}(\tilde{q}) = \frac{1}{2\sigma}[v(\tilde{q}) - 1], \qquad (9)$$

where $v(\tilde{q}) \equiv v(x, y) = \alpha x + \beta y = a^2 U(\phi)$.

Using these convenient properties Halliwel, Marugán and Garay [42] demonstrated that the WDW equation (derived from our Eq.(8),(9)) has an exact general solution which is a linear combination of the products of Airy functions. Also, by applying the Hartle-Hawking condition [38], [39] (the H-H condition, in what follows) and the saddle point method to the path integral over the complex lapse parameter (under the gauge condition $\dot{N} = 0$) and the real minisuperspace coordinates $x$ and $y$, they found the particular solutions to the WDW equation. For different choices of the contours of integration over $N$ and for different values of parameters $\alpha$ and $\beta$, one obtains integral representations of different special functions.

The non-commutative model of the closed universe can now be realized by application of the Moyal deformation [49] to the WDW equation, what is a procedure well known from the case of the non-commutative quantum mechanics. As the minisuperspace coordinates depend on time only, the non-commutativity of space-time is trivially realized by the suitable non-commutative cosmology. As the minisuperspace model of the universe is a quantum-mechanical model with an infinite speed of interaction, and not the quantum field theory of the universe, it does not violate the microcausality.

In the case of our model we suggest the canonical non-commutativity of linear minisuperspace coordinates $\hat{\tilde{q}}^A = (\hat{x}, \hat{y})$ determined by commutation relations:

$$\left[\hat{\tilde{q}}^A, \hat{\tilde{q}}^B\right] = i\theta^{AB} \equiv i\theta\varepsilon^{AB}, \quad \left[\hat{\tilde{q}}^A, \hat{\tilde{p}}_B\right] = i\delta^A_B, \quad \left[\hat{\tilde{p}}_A, \hat{\tilde{p}}_B\right] = 0, \tag{10}$$

where the non-commutativity real parameter $\theta = const$, $\hat{\tilde{p}}_A \equiv (\hat{p}_x, \hat{p}_y)$ - are linear minisuperspace momenta and $\varepsilon^{AB} = -\varepsilon^{BA}$ ($\varepsilon^{01}=1$). The $\theta$ - deformation of the minisuperspace as determined by (10) implies the non-commutativity of the minisuperspace functions that is introduced via the Moyal (star) product [49]:

$$\psi_1(\tilde{q}^A_C) \star \psi_2(\tilde{q}^B_C) = \psi_1(\tilde{q}^A_C)\exp\left(\frac{i\theta}{2}\varepsilon^{CD}\overleftarrow{\tilde{\partial}}_C \overrightarrow{\tilde{\partial}}_D\right)\psi_2(\tilde{q}^B_C) =$$

$$= \psi_1(\tilde{q}^A_C)\psi_2(\tilde{q}^B_C) + \frac{i\theta}{2}\varepsilon^{CD}\left[\tilde{\partial}_C\psi_1(\tilde{q}^A_C)\right]\tilde{\partial}_D\psi_2(\tilde{q}^B_C) + O(\theta^2), \tag{11}$$

where $\tilde{\partial}_A \equiv \frac{\partial}{\partial \tilde{q}^A_C}$ and the commutative minisuperspace coordinates $\tilde{q}^A_C = (x_C, y_C)$ are the Weyl symbols of the operators $\hat{\tilde{q}}^A = (\hat{x}, \hat{y})$. The notation here is suggested because these symbols match exactly with the canonical minisuperspace coordinates as defined by the following formula (18), what may be seen directly from the properties of the Moyal product (11) [2], [44]. Our non-commutative WDW (NWDW) equation now looks like:

$$\left[4\left(-\hat{p}_x^2 + \hat{p}_y^2\right) + v(x_C, y_C) - 1\right] \star \psi(x_C, y_C) = 0 \tag{12}$$

and represents the Moyal deformed version of the WDW equation. Applying the properties of the Moyal product to the potential term $v(x_C, y_C)$ in (12) we obtain:

$$v(x_C, y_C) \star \psi(x_C, y_C) = v\left(x_C + \frac{i\theta}{2}\partial_{y_C}, y_C - \frac{i\theta}{2}\partial_{x_C}\right)\psi(x_C, y_C) =$$

$$= v\left(\hat{x}_C - \frac{\theta}{2}\hat{p}_y, \hat{y}_C + \frac{\theta}{2}\hat{p}_x\right)\psi(x_C, y_C) = v(\hat{x}, \hat{y})\psi(x_C, y_C) \tag{13}$$

and our NWDW equation can now be written as:

$$\left[4\left(\partial^2_{x_C} - \partial^2_{y_C}\right) + \frac{i\theta}{2}\left(\alpha\partial_{y_C} - \beta\partial_{x_C}\right) + \alpha x_C + \beta y_C - 1\right]\psi(x_C, y_C) = 0 \tag{14}$$

with the general non-commutative solution of the form:

$$\psi_\theta(x_C, y_C) = \left\{ A' \text{Ai}\left[\frac{1+\mu'-\alpha\, x_C - \theta^2 \beta^2/64}{(2\alpha)^{2/3}}\right] + B' \text{Bi}\left[\frac{1+\mu'-\alpha\, x_C - \theta^2 \beta^2/64}{(2\alpha)^{2/3}}\right] \right\} \times$$

$$\times \left\{ C' \text{Ai}\left[\frac{\mu'+\beta\, y_C - \theta^2 \alpha^2/64}{(2\beta)^{2/3}}\right] + D' \text{Bi}\left[\frac{\mu'+\beta\, y_C - \theta^2 \alpha^2/64}{(2\beta)^{2/3}}\right] \right\} \exp\left[\frac{i\theta}{16}(\beta x_C + \alpha y_C)\right],$$

(15)

where $A'$, $B'$, $C'$, $D'$ and $\mu'$ are arbitrary constants. For $\theta = 0$ this non-commutative solution reduces to the general solution [42] of the common WDW equation.

As the metric of the space-time (1) of the closed model universe is compact (after applying the Wick rotation of time) it is possible to find the particular solutions of the NWDW equation by applying the H-H condition and using the $\theta$-modified method of path integrals. If, in a sense of the modified Poisson brackets, we assume the non-commutativity of minisuperspace coordinates $\tilde{q}^A = (x, y)$ on the classical level, we then have to define the $\theta$-deformed Poisson algebra:

$$\{\tilde{q}^A, \tilde{q}^B\} = \theta \varepsilon^{AB}, \qquad \{\tilde{q}^A, \tilde{p}_B\} = \delta^A_B, \qquad \{\tilde{p}_A, \tilde{p}_B\} = 0,$$

(16)

Classical non-commutative algebra (16) is related to the commutation relations (10) by the Dirac quantization procedure:

$$\{\ ,\ \} \to \frac{1}{i}[\ ,\ ].$$

(17)

By applying the linear transformations

$$\tilde{q}^A = \tilde{q}^A_C - \frac{\theta \varepsilon^{AB}}{2} \tilde{p}_{CB},$$

(18)

where is $\tilde{p}_{CB} = \tilde{p}_B$ because of the third commutation relation in (16), one obtains the standard Poisson algebra:

$$\{\tilde{q}_C^A, \tilde{q}_C^B\} = 0, \qquad \{\tilde{q}_C^A, \tilde{p}_B\} = \delta^A_B, \qquad \{\tilde{p}_A, \tilde{p}_B\} = 0,$$

(19)

and the canonical Hamiltonian of our model becomes:

$$\tilde{H}^\theta \equiv N\tilde{H}_\theta = \frac{N}{2}\left[4\left(-p_x^2+p_y^2\right)-\frac{\theta}{2}\left(\alpha p_y-\beta p_x\right)+\alpha x_C+\beta y_C-1\right]\approx 0. \tag{20}$$

By applying the Legendre transformation and the Wick rotation of time to the Hamiltonian (20) the non-commutative Euclidean action:

$$\tilde{I}_\theta \stackrel{t\to-i\tau}{=} -i\tilde{S}_\theta = \frac{1}{2}\int_0^1 d\tau\; N\,\Big[\frac{1}{4N^2}\left(-\dot{x}_C^2+\dot{y}_C^2\right)-\frac{i\theta}{8N}\left(\beta\dot{x}_C+\alpha\dot{y}_C\right)-$$
$$-\frac{\theta^2}{64}\left(\alpha^2-\beta^2\right)+\alpha x_C+\beta y_C-1\Big], \tag{21}$$

is obtained (the dot over the symbol signifies derivation over $\tau$), and equations of motion become:

$$\ddot{x}_C=-2\alpha N^2 \text{ and } \ddot{y}_C=2\beta N^2 \tag{22}$$

and their solutions:

$$\bar{x}_C(\tau)=-\alpha N^2\tau^2+\left(x_C''-x_C'+\alpha N^2\right)\tau+x_C',$$
$$\bar{y}_C(\tau)=\beta N^2\tau^2+\left(y_C''-y_C'-\beta N^2\right)\tau+y_C', \tag{23}$$

where $x_C''\equiv \bar{x}_C(1)$, $y_C''\equiv \bar{y}_C(1)$, $x_C'\equiv \bar{x}_C(0)$, $y_C'\equiv \bar{y}_C(0)$, henceforth follow. In Eq. (21)

$$\tilde{S}_\theta = \int_0^1 dt\,\tilde{L}_\theta = \frac{1}{2}\int_0^1 dt\; N\,\left\{\frac{1}{4N^2}\left[-\left(\frac{dx_C}{dt}\right)^2+\left(\frac{dy_C}{dt}\right)^2\right]+\frac{\theta}{8N}\left(\beta\frac{dx_C}{dt}+\alpha\frac{dy_C}{dt}\right)+\right.$$
$$\left.+\frac{\theta^2}{64}\left(\alpha^2-\beta^2\right)-\alpha x_C-\beta y_C+1\right\} \tag{24}$$

is the Lorentz action. Inserting these solutions into the Euclidean action we obtain the non-commutative Hamilton-Jacobi action:

$$\bar{\tilde{I}}_\theta\left(x_C'', y_C'', N \mid x_C', y_C', 0\right) = \frac{1}{8N}\left[-(x_C'' - x_C')^2 + (y_C'' - y_C')^2\right] + \frac{N^3}{24}(\alpha^2 - \beta^2) -$$

$$-\frac{N}{4}\left[2 - \alpha(x_C'' + x_C') - \beta(y_C'' + y_C') + \frac{\theta^2}{32}(\alpha^2 - \beta^2)\right] -$$

$$-\frac{i\theta}{16}[\beta(x_C'' - x_C') + \alpha(y_C'' - y_C')]. \tag{25}$$

Under the H-H condition and the gauge condition $\dot{N} = 0$, the non-commutative quantum-mechanical propagator $G_{\theta E}^{HH}\left(\tilde{q}_C''^A, N \mid 0, 0\right)$ can now be exactly found from the Pauli formula [52], [53]

$$G_{\theta E}^{HH}\left(\tilde{q}_C''^A, N \mid 0, 0\right) = \frac{1}{(2\pi)^{n/2}} \sqrt{-\det\left(\frac{\partial^2 \bar{\tilde{I}}_\theta}{\partial \tilde{q}_C''^A \partial \tilde{q}_C'^A}\right)} \exp\left(-\bar{\tilde{I}}_\theta\right)\Bigg|_{\tilde{q}_C'^A = 0}, \tag{26}$$

where $n = 2$ is the dimension of our minisuperspace.
The propagator is:

$$G_{\theta E}^{HH}\left(x_C'', y_C'', N \mid 0, 0, 0\right) = \frac{1}{8\pi N} \exp\left\{-\frac{1}{4}\left\{\frac{-x_C''^2 + y_C''^2}{2N} + \frac{N^3}{6}(\alpha^2 - \beta^2) - \right.\right.$$

$$\left.\left. -N\left[2 - \alpha x_C'' - \beta y_C'' + \frac{\theta^2}{32}(\alpha^2 - \beta^2)\right]\right\}\right\} \times \exp\left[\frac{i\theta}{16}(\beta x_C'' + \alpha y_C'')\right]. \tag{27}$$

When $\alpha = \beta$ the propagator (27) does not contain the $\theta^2$ dependence so that integration over the complex lapse parameter by applying the method of fastest descent yields the particular solutions of the NWDW equation [41]:

$$\psi_{\theta NB}(x_C'', y_C'') = \int dN \, G_{\theta E}^{HH}(x_C'', y_C'', N \mid 0, 0, 0) = \exp\left[\frac{i\theta\alpha}{16}(x_C'' + y_C'')\right] \psi_{NB}(x_C'', y_C''), \tag{28}$$

where $\psi_{NB}(x_C'', y_C'')$ are all particular solutions of the WDW equation, as obtained by Halliwell et al. [42]. Although the particular solutions (28) depend on $\theta$, the corresponding general solution for the $\alpha = \beta$ case contain the $\theta^2$ dependence:

$$\psi_\theta(x_C, y_C) = \left\{ A' \operatorname{Ai}\left[\frac{1+\mu'-\alpha\, x_C - \theta^2 \alpha^2/64}{(2\alpha)^{2/3}}\right] + B' \operatorname{Bi}\left[\frac{1+\mu'-\alpha\, x_C - \theta^2 \alpha^2/64}{(2\alpha)^{2/3}}\right] \right\} \times$$

$$\times \left\{ C' \operatorname{Ai}\left[\frac{\mu'+\alpha\, y_C - \theta^2 \alpha^2/64}{(2\alpha)^{2/3}}\right] + D' \operatorname{Bi}\left[\frac{\mu'+\alpha\, y_C - \theta^2 \alpha^2/64}{(2\alpha)^{2/3}}\right] \right\} \exp\left[\frac{i\theta\alpha}{16}(x_C + y_C)\right].$$

(29)

This $\theta^2$ dependence in the general solution (29) is only illusory, as we will see from what follows. Comparison of the general solution and all particular solutions for $\alpha = \beta$ shows that under the H-H condition the constant $\mu'$ that figures in (29) equals

$$\mu' = \mu + \frac{\theta^2 \alpha^2}{64} \qquad (30)$$

where $\mu$ is the undetermined constant of the general solution of the WDW equation (derived from our Eq.(8), (9)) and other constants in general solution (29) have the same values as in the commutative (standard) case [42].

So, for $\alpha = \beta$ the $\theta$- phase factor which appears in (28) and (29) influences the quantum-mechanical interference between different configurations of the universe.

Let us consider now the case $\alpha \neq \beta$, when in (15), (25) and (27) there appears the term that contains $\theta^2$. Herewith we shall apply the H-H condition and the semiclassical approximation. In the standard case, when $\theta = 0$ and $\alpha > |\beta|$, under the H-H condition with the regularity condition $s^2 \equiv -x_C''^2 + y_C''^2 < 0$, Halliwell et. al. applied the contour analysis and obtained the semiclassical wave functions [42]:

$$\psi_{a/b}^\pm(x_C'', y_C'') \propto \exp\left[-\overline{I}_{a/b}^\pm\left(x_C'', y_C'', N_{a/b}^\pm \Big| 0,0,0\right)\right], \qquad (301)$$

as well as the sum of the wave functions of this form, where the Hamilton-Jacobi action (see (25)):

$$\overline{I}_{a/b}^\pm \equiv \overline{I}_\theta\left(x_C'', y_C'', N_{a/b}^\pm \Big| 0,0,0\right)\Big|_{\theta=0} \qquad (302)$$

has the saddle points of the form:

$$N^{\pm}_{a/b} = \frac{\pm 1}{\sqrt{\alpha^2 - \beta^2}} \left[ -I \pm \sqrt{I^2 + (\alpha^2 - \beta^2)s^2} \right]^{1/2}, \quad (I \equiv \alpha x''_C + \beta y''_C - 2) \tag{303}$$

(in Eq. (303) the superscript "$\pm$" on $N$ corresponds to $\pm 1$ in front of the bracket on the right hand side of Eq. (303), while the subscript "a/b" corresponds to $\pm 1$ in front of the square root in the bracket on the same side of Eq. (303)).

For $\alpha > 0$, $\beta = 0$, and $y''^2_C = 0$, and under the gauge condition:

$$N'^{-}_a = \frac{-1}{\alpha}\left(-I' + \sqrt{I'^2 + \alpha^2 s^2}\right)^{1/2}, \quad (I' = \alpha x''_C - 2) \tag{304}$$

the corresponding Hamilton-Jacobi action:

$$\widetilde{I}'^{-}_a = \frac{-1}{6\alpha}\left(-I' + \sqrt{I'^2 + \alpha^2 s^2}\right)^{1/2}\left(2I' + \sqrt{I'^2 + \alpha^2 s^2}\right) \equiv \widetilde{I}_0\left(x''_C, 0, N'^{-}_a \big| 0,0,0\right)\Big|_{\theta=0} \tag{305}$$

is real, for $I' = \alpha x''_C - 2 < 0$ and $I'^2 + \alpha^2 s^2 > 0$.

Action (305) determines the semiclassical wave function:

$$\psi'^{-}_a(x''_C) \propto \exp\left[-\widetilde{I}'^{-}_a\left(x''_C, 0, N'^{-}_a \big| 0,0,0\right)\right] \tag{306}$$

which for $\alpha x''_C \ll 1$ corresponds to the Hartle-Hawking (no-boundary) ground state of the closed universe [38], [39].

Starting with this, we choose for our consideration of the noncommutative model of the closed universe with $\alpha > |\beta|$ the following gauge conditions:

$$N^{\pm}_{\theta a} = \frac{\pm 1}{\sqrt{\alpha^2 - \beta^2}}\left[-I_\theta + \sqrt{I_\theta^2 + (\alpha^2 - \beta^2)s^2}\right]^{1/2}, \quad \left(I_\theta \equiv \alpha x''_C + \beta y''_C - 2\left[1 + \frac{\theta^2(\alpha^2 - \beta^2)}{64}\right]\right) \tag{307}$$

which are also the saddle points of action (25). From (307), if:

$$I_\theta < 0 \Leftrightarrow a''^2_C U(\phi''_C) < 2\left[1 + \frac{\theta^2(\alpha^2 - \beta^2)}{64}\right] \quad \text{and} \quad I_\theta^2 + (\alpha^2 - \beta^2)s^2 > 0 \tag{3071}$$

then $|I_\theta| > |I|$, and for the gauge condition $N^{-}_{\theta a}$ the following relation hold:

$$\left(\widetilde{I}^{R-}_{\theta a}, \widetilde{I}^{-}_a \in \mathbb{R}\right) \; \widetilde{I}^{R-}_{\theta a} > \widetilde{I}^{-}_a \Rightarrow \psi^{R-}_{\theta a} < \psi^{-}_a \tag{308}$$

Here:

$$\overline{\overline{I}}_{\theta a}^{R-} = \text{Re}\left[\overline{\overline{I}}_{\theta a}^{-} \equiv \overline{I}_{\theta}\left(x_C'', y_C'', N_{\theta a}^{-} \middle| 0,0,0\right)\right]$$

and the total semiclassical wave function $\psi_{\theta a}^{-}$ takes the form:

$$\psi_{\theta a}^{-}(x_C'', y_C'') \propto \psi_{\theta a}^{R-} \exp\left[\frac{i\theta}{16}(\beta x_C'' + \alpha y_C'')\right], \left(\psi_{\theta a}^{R-} = \exp\left(-\overline{\overline{I}}_{\theta a}^{R-}\right)\right). \quad (309)$$

The $\theta$ - phase factor that appears in (309) does not vary with $x_C''$ and $y_C''$ much more rapidly than $\overline{\overline{I}}_{\theta a}^{R-}$ [41], so that its contribution to interference phenomena is non-trivial.

On the basis of (308) we conclude that for $\alpha > |\beta|$ and the gauge condition $N_{\theta a}^{-}$ the noncommutativity via the $\theta^2$ dependence decreases the semiclassical tunneling amplitude from nothing to the observable closed universe with stable matter potential.

Under the same conditions (3071), but for the gauge condition $N_{\theta a}^{+}$ the following holds:

$$\left(\overline{\overline{I}}_{\theta a}^{R+}, \overline{\overline{I}}_{a}^{+} \in \text{R}\right) \overline{\overline{I}}_{\theta a}^{R+} < \overline{\overline{I}}_{a}^{+} \Rightarrow \psi_{\theta a}^{R+} > \psi_{a}^{+}, \quad (3010)$$

where

$$\overline{\overline{I}}_{\theta a}^{R+} = \text{Re}\left[\overline{\overline{I}}_{\theta a}^{R+} \equiv \overline{I}_{\theta}\left(x_C'', y_C'', N_{\theta a}^{R+} \middle| 0,0,0\right)\right]$$

and the total semiclassical wave function $\psi_{\theta a}^{+}$ takes the form:

$$\psi_{\theta a}^{+}(x_C'', y_C'') \propto \psi_{\theta a}^{R+} \exp\left[\frac{i\theta}{16}(\beta x_C'' + \alpha y_C'')\right], \left(\psi_{\theta a}^{R+} = \exp\left(-\overline{\overline{I}}_{\theta a}^{R+}\right)\right). \quad (3011)$$

The $\theta$- phase factor in (3011) again has non-trivial influence only on interference effects.

From (3010) there follows that, opposite to the previous case, for $\alpha > |\beta|$ and the gauge condition $N_{\theta a}^{+}$, the noncommutativity increases the tunneling amplitude.

Also, from (3071) we see that the noncommutative $\theta^2$ term extends the classically forbidden, i.e. the quantum-mechanically allowed region of the minisuperspace, thereby increasing the difference between the contributions of the semiclassical (no-boundary) tunneling amplitudes $\psi_{\theta a}^{-}$ and $\psi_{\theta a}^{+}$.

This difference is greatest for $\alpha > 0$ and $\beta = 0$, what means that noncommutativity prefers the creation of the closed universe rather by $\psi_{\theta a}^{+}$ than by $\psi_{\theta a}^{-}$.

For $|\alpha| < \beta$ the gauge conditions (307) are real, if:

$$I_\theta \equiv \alpha x_C'' + \beta y_C'' - 2\left(1 - \frac{\theta^2 |\beta^2 - \alpha^2|}{64}\right) > 0 \Leftrightarrow a_C''^2 U(\phi_C'') > 2\left(1 - \frac{\theta^2 |\beta^2 - \alpha^2|}{64}\right) \quad (3012)$$

The corresponding semiclassical wave functions are then of the form:

$$\psi_{\theta a}^{\pm}(x_C'', y_C'')\bigg|_{\alpha < |\beta|} \propto \psi_{\theta a}^{R\pm} \exp\left[\frac{i\theta}{16}(\beta x_C'' + \alpha y_C'')\right]\bigg|_{\alpha < |\beta|}, \quad \left(\psi_{\theta a}^{R\pm} = \exp\left(-\widetilde{I}_{\theta a}^{R\pm}\right)\right) \quad (3013)$$

where the the $\theta$- phase factor again takes part in quantum interference. Minding that on the grounds of (3012) in this case the large cosmic scale structure of our model universe is quantum-mechanically allowed, or classically forbidden, the wave functions (3013) in this region do not decohere. This means that there is no creation of the classical closed universe with unstable matter potential, to what the contribution of noncommutativity is negative, and decreases the probability of creation.

These considerations tell us that noncommutativity of minisuperspace model of the closed universe prefers creation of the classical universe with stable matter potential only, what follows from Eq. (3010), (3011) and (3013).

Let us consider now the noncommutative geometry of our classical model of the universe. If we employ finite diffeomorphism $\widetilde{q}_C^{\ A} \equiv (x_C, y_C) \rightarrow q_C^{\ A} \equiv (a_C, \phi_C)$ which is defined by

$$x_C = a_C^2 \operatorname{ch} 2\phi_C, \quad y_C = a_C^2 \operatorname{sh} 2\phi_C \quad (34)$$

to the noncommutative Lorentz lagrangian represented by (24) we obtain the noncommutative Lorentz lagrangian in the form:

$$L_\theta = \frac{N}{2}\Bigg\{\frac{1}{N^2}\left[-a_C^2\left(\frac{da_C}{dt}\right)^2 + a_C^4\left(\frac{d\phi_C}{dt}\right)^2\right] - a_C^2(\alpha \operatorname{ch}2\phi_C + \beta \operatorname{sh}2\phi_C) + 1 +$$

$$+ \frac{\theta}{4N}\left[a_C(\alpha \operatorname{sh}2\phi_C + \beta \operatorname{ch}2\phi_C)\frac{da_C}{dt} + a_C^2(\alpha \operatorname{ch}2\phi_C + \beta \operatorname{sh}2\phi_C)\frac{d\phi_C}{dt}\right] + \frac{\theta^2}{64}(\alpha^2 - \beta^2)\Bigg\}.$$

(35)

From (24), (34) and (35) there follows the finite smooth transformation $\tilde{p}_A \equiv (p_x, p_y) \to p_{CA} \equiv (p_{a_C}, p_{\phi_C})$ defined by the following relations

$$p_x = \frac{1}{2a_C^2}(p_{a_C} a_C \text{ch} 2\phi_C - p_{\phi_C} \text{sh} 2\phi_C), \qquad p_y = \frac{1}{2a_C^2}(-p_{a_C} a_C \text{sh} 2\phi_C + p_{\phi_C} \text{ch} 2\phi_C). \qquad (36)$$

Transformations (34) and (36) conserve the standard Poisson algebra (19) i.e. their application to this algebra yields the following Poisson algebra:

$$\{q_C^A, q_C^B\} = 0, \qquad \{q_C^A, p_{CB}\} = \delta_B^A, \qquad \{p_{CA}, p_{CB}\} = 0, \qquad (37)$$

and the Legandre transformation of (35) defines the noncommutative Hamiltonian

$$H^\theta \equiv N H_\theta = \frac{N}{2}\left\{-\frac{p_{a_C}^2}{a_C^2} + \frac{p_{\phi_C}^2}{a_C^4} + a_C^2(\alpha \text{ch} 2\phi_C + \beta \text{sh} 2\phi_C) - 1 - \right.$$

$$\left. -\frac{\theta}{4a_C^2}\left[-p_{a_C} a_C(\beta \text{ch} 2\phi_C + \alpha \text{sh} 2\phi_C) + p_{\phi_C}(\alpha \text{ch} 2\phi_C + \beta \text{sh} 2\phi_C)\right]\right\} \approx 0. \qquad (38)$$

Also, the noncommutative Hamiltonian (38) may be obtained by applying the transformation (36) to the Hamiltonian (20).

Earlier we showed that algebra (19) may be obtained from the $\theta$-deformed Poisson algebra (16) by using linear transformations (18). Analogously, if we use a natural substitution of the noncommutative minisuperspace coordinates $\tilde{q}^A \equiv (x, y) \to q^A \equiv (a, \phi)$ defined by the relations:

$$x = a^2 \text{ch} 2\phi, \quad y = a^2 \text{sh} 2\phi, \qquad (39)$$

as well as the substitution of momenta $\tilde{p}_A \equiv (p_x, p_y) \to p_A \equiv (p_a, p_\phi)$ defined by terms

$$p_x = \frac{1}{2a^2}(p_a a \text{ ch} 2\phi - p_\phi \text{sh} 2\phi), \quad p_y = \frac{1}{2a^2}(-p_a a \text{ sh} 2\phi + p_\phi \text{ ch} 2\phi) \qquad (40)$$

in (16) one gets the $\theta$-deformed Poisson algebra:

$$\{a, \phi\} = \frac{\theta}{4a^3}, \{a, p_a\} = 1 + \frac{\theta p_\phi}{2a^4}, \{\phi, p_\phi\} = 1 - \frac{\theta p_\phi}{2a^4}, \{\phi, p_a\} = -\frac{\theta p_a}{4a^4},$$

$$\{a, p_\phi\} = \frac{\theta p_a}{2a^2}, \{p_a, p_\phi\} = \theta\left(\frac{p_a^2}{2a^3} - \frac{p_\phi^2}{a^5}\right). \tag{41}$$

Although relations (39) are satisfied, as well as the equalities $\tilde{p}_{C_A} = \tilde{p}_A$, the equality $p_A = p_{C_A}$ is not satisfied because of the obvious difference of algebras (37) and (41). The theta-deformed Poisson algebra (41) may be written as:

$$\{q^A, q^B\} = \frac{\theta \varepsilon^{AB}}{\sqrt{-G^*(q)}}, \{q^A, p_B\} = \delta^A_B + \frac{\theta \varepsilon^{AC}}{\sqrt{-G^*(q)}} \Gamma^{*D}_{CB} p_D,$$

$$\{p_A, p_B\} = \frac{\theta \varepsilon^{CD}}{\sqrt{-G^*(q)}} \Gamma^{*E}_{CA} \Gamma^{*F}_{DB} p_E p_F, \tag{42}$$

where the $G^*(q)$-determinant of the redefined minisuperspace metric, which is defined as $G^*_{AB}(q) = \frac{4}{\sigma} G_{AB}(q)$ (see the first formula in (4)), $\varepsilon^{AB} = -\varepsilon^{BA}$ ($\varepsilon^{01} = 1$) and $\Gamma^{*A}_{BC} = \frac{G^{*AD}(q)}{2}\left(G^*_{DB,C} + G^*_{DC,B} - G^*_{BC,D}\right)\left(G^*_{AB,C} \equiv \frac{\partial G^*_{AB}}{\partial q^C}\right)$, is the Christoffel symbol of the redefined minisuperspace metric. Therefore, diffeomorphism (39) of the noncommutative minisuperspace and transformations of momenta (40) nontrivially change $\theta$-deformed algebra (16) to algebra (42). From (34), (36), (39) and (40), and using the linear transformations (18), we obtain the nonlinear transformations:

$$a = \left[a_C^4 - \frac{\theta p_{\phi_C}}{2} + \frac{\theta^2}{4}\Pi(q_C, p_C)\right]^{1/4} = a_C - \frac{\theta p_{\phi_C}}{8a_C^3} - \frac{\theta^2}{64 a_C^7}\left(\frac{p_{\phi_C}^2}{2} + a_C^2 p_{a_C}^2\right) + O(\theta^3),$$

$$\phi = \phi_C + \frac{1}{4}\ln\left|\frac{1 - \frac{\theta}{4a_C^4}(p_{\phi_C} - a_C p_{a_C})}{1 - \frac{\theta}{4a_C^4}(p_{\phi_C} + a_C p_{a_C})}\right| = \phi_C + \frac{\theta p_{a_C}}{8 a_C^3} + \frac{\theta^2 p_{a_C} p_{\phi_C}}{32 a_C^7} + O(\theta^3),$$

$$p_a = \frac{a_C p_{a_C}}{\left[a_C^4 - \frac{\theta p_{\phi_C}}{2} + \frac{\theta^4}{4}\Pi(q_C, p_C)\right]^{1/4}} = p_{a_C} + \frac{\theta p_{a_C} p_{\phi_C}}{8 a_C^4} + \frac{\theta^2 p_{a_C}}{64 a_C^8}\left(a_C^2 p_{a_C}^2 + 2 p_{\phi_C}^2\right) + O(\theta^3),$$

$$p_\phi = p_{\phi_C} - \theta \Pi(q_C, p_C), \tag{43}$$

where $\Pi(q_C, p_C) \equiv G^{*AB}(q_C) p_{CA} p_{CB}$. By applying these nonlinear transformations to (42) one obtains the standard Poisson algebra (37). This is possible because of the invariance of the mathematical object $\Pi(q_C, p_C)$ to transformations (36) and (40) i.e.:

$$\Pi(q,p) = G^{*AB}(q) p_A p_B = \frac{\sigma}{4} \tilde{G}^{AB} \tilde{p}_A \tilde{p}_B = G^{*AB}(q_C) p_{CA} p_{CB} = \Pi(q_C, p_C). \qquad (44)$$

Also, the mathematical object (44) is represented in the form of quadratic terms of momenta in noncommutative Hamiltonians (20) and (38), so that these parts of the Hamiltonians are invariant to application of (18) and (43), respectively. Therefore, if $F(q^A, p_A), G(q^B, p_B) \in C^\infty(M_p^\theta) \left( M_p^\theta \equiv \left( M_p, \{ , \}_\theta \right) \right)$ are smooth functions on the noncommutative phase minisuperspace $M_p^\theta$ with local minisuperspace coordinates $q^A$ and momenta $p_A$, then one may define the following $\theta$-modified Poisson bracket:

$$\{F(q^A, p_A), G(q^B, p_B)\}_\theta = \left(\frac{\partial F}{\partial q^A}\right) \{q^A, q^B\} \frac{\partial G}{\partial q^A} + \left(\frac{\partial F}{\partial q^A}\right) \{q^A, p_B\} \frac{\partial G}{\partial p_B} +$$

$$+ \left(\frac{\partial F}{\partial p_A}\right) \{p_A, q^B\} \frac{\partial G}{\partial q^B} + \left(\frac{\partial F}{\partial p_A}\right) \{p_A, p_B\} \frac{\partial G}{\partial p_B}, \qquad (45)$$

where $\{q^A, q^B\} \neq 0, \{p_A, p_B\} \neq 0$ and $\{q^A, p_B\} = -\{p_B, q^A\}$ are determined by algebra (42). The noncommutative geometry and noncommutative Hamiltonian dynamics of the classical model of the closed universe is now completely defined by the construction of the modified Poisson brackets (45).

Before we start considering the noncommutative geometry of the minisuperspace of the quantum model of the closed universe it is important to note that the WDW equation (5) of the standard (commutative) nD minisuperspace model may be obtained from the action:

$$S_n[\mathbf{G}_{AB}(q), \Psi, \overline{\Psi}] = -\int_M d^n q \sqrt{-\mathbf{G}} \left\{ \frac{1}{2} \mathbf{G}^{AB} \nabla_A \overline{\Psi} \nabla_B \Psi + \overline{\Psi} [\xi R + \mathbf{U}(q)] \Psi \right\} =$$

$$= -\int_M d^n q \sqrt{-\mathbf{G}} \, \overline{\Psi} \left[ -\frac{1}{2} \Delta + \xi R + \mathbf{U}(q) \right] \Psi = -\int_M d^n q \sqrt{-\mathbf{G}} \, \overline{\Psi} H \Psi, \qquad (46)$$

by the Lagrange variation of $\overline{\Psi}(q^A)$. Physically, action (46) is related to the expectation value of the energy of the Universe, which is invariant to the conformal transformation (Weyl rescaling) for the fixed value of the ordering parameter $\xi$. In the case when nD minisuperspace $M^\bullet$ possesses Weyl geometry [54] the action

$$\overset{w'}{S'}_n\left[\mathbf{G}'_{AB}(q),\Psi',\overline{\Psi}',w'_A\right] = -\int_{M^\bullet} d^n q \sqrt{-\mathbf{G}'}\left\{\frac{1}{2}\mathbf{G}'^{AB}\overset{\bullet}{\nabla}_A\overline{\Psi}'\overset{\bullet}{\nabla}_B\Psi' + \overline{\Psi}'\left[\xi R'(\overset{\bullet}{\Gamma'}) + U'(q)\right]\Psi'\right\},$$

$$\left(R'(\overset{\bullet}{\Gamma'}) = R'(\overset{\bullet}{\overset{w'}{\Gamma'}})\right),$$

(47)

may be constructed, where $\overset{\bullet}{\nabla}_A \equiv \nabla_A - \frac{1}{2}\rho w'_A$ is the cocovariant derivative ($\rho$- is the Weyl weight and $w'_A$-is the Weyl vector),

$$\overset{\bullet}{\Gamma'}{}^B_{CA} \equiv \overset{w'}{\Gamma'}{}^B_{CA} - \frac{1}{2}\rho\delta^B_C w'_A\left(\overset{w'}{\Gamma'}{}^B_{CA} = \Gamma'{}^B_{CA} - \frac{1}{2}\left(\delta^B_C w'_A + \delta^B_A w'_C - \mathbf{G}'_{CA}w'^B\right)\right)$$ is the Weyl connection and

$$R'(\overset{\bullet}{\Gamma'}) = R'(\overset{w'}{\Gamma'}) = R'(\Gamma') + (n-1)(2-n)\frac{w'^2}{4} + (n-1)\nabla(\Gamma')w' =$$

$$= R'(\Gamma') - (n-1)(2-n)\frac{w'^2}{4} + (n-1)\nabla(\overset{w'}{\Gamma'})w',$$

$$\left(\nabla(\Gamma')w' = \partial_A w'^A + \Gamma'^A_{BA} w'^B; \nabla(\overset{w'}{\Gamma'})w' = \partial_A w'^A + \overset{w'}{\Gamma'}{}^A_{BA} w'^B\right) \quad (48)$$

is the scalar Weyl curvature. The action (47) is invariant on Weyl rescaling [50]:

$$\mathbf{G}_{AB}(q) = \Omega^2(q)\mathbf{G}'_{AB}(q),\ \Psi(q^A) = \Omega^{1-\frac{n}{2}}(q)\Psi'(q^A),\ \overline{\Psi}(q^A) = \Omega^{1-\frac{n}{2}}(q)\overline{\Psi}'(q^A),\ U(q) = \Omega^{-2}(q)U'(q),$$

(49)

where $n$ is the dimension of the minisuperspace and the rescaling

$$w_A = w'_A + 2\Omega^{-1}\partial_A\Omega, \qquad (49_1)$$

for any value of the ordering parameter $\xi$ due to the validity of the following relation:

$$R(\overset{\bullet}{\Gamma}) = R(\overset{w}{\Gamma}) = R(\overset{\bullet}{\overset{w'}{\Gamma'}}) = R(\overset{\bullet}{\Gamma'}) = \mathbf{G}^{AB} R_{AB}(\overset{\bullet}{\Gamma'}) = \mathbf{G}^{AB} R'_{AB}(\overset{\bullet}{\Gamma'}) = \Omega^{-2}(q) R'(\overset{\bullet}{\Gamma'}). \quad (50)$$

When minisuperspace is (conformally) flat, then action (47) acquires the following form:

$$\overset{w'}{S'}_n\left[\mathbf{G}'_{AB}(q),\Psi',\overline{\Psi}',w'_A\right] = -\int_{M^\bullet} d^n q\sqrt{-\mathbf{G}'}\left[\frac{1}{2}\mathbf{G}'^{AB}\overset{\bullet}{\partial}_A\overline{\Psi}'\overset{\bullet}{\partial}_B\Psi' + \overline{\Psi}' U'(q)\Psi'\right]$$

(51)

and by applying the Weyl rescaling (49) and the rescaling

$$w_A = w'_A + 2\Omega^{-1} \partial_A \Omega = 0 \tag{52}$$

one obtains the action:

$$S_n[\mathbf{G}_{AB}(q), \Psi, \overline{\Psi}, w_A = 0] = -\int_M d^n q \sqrt{-\mathbf{G}} \left[ \frac{1}{2} \mathbf{G}^{AB} \partial_A \overline{\Psi} \partial_B \Psi + \overline{\Psi} \mathbf{U}(q) \Psi \right] \tag{53}$$

which is same as action (46) for which the Riemann scalar curvature vanishes. The rescaling (52) represents the gauge-fixing condition, because its choice violates the Weyl rescaling symmetry of the flat minisuperspace and determines the value of the ordering parameter $\xi$ for which Eq. (5) is still conformally invariant but this value is not important because the second term in Eq. (5) is equal to zero. This simplification leads to the following considerations of our model, with appropriate adaptation in the noncommutative case, the minimal realization of which follows from the Poincare-Birkhoff-Witt theorem by replacing the ordinary product with the Moyal product leaving the minisuperspace coordinates of the standard (commutative) minisuperspace as Weyl symbols [44]. Actually, in our case the analogue of the action:

$$S_2[\widetilde{G}_{AB}, \psi, \overline{\psi}] = -\int_M d^2 \widetilde{q} \sqrt{-\widetilde{G}} \left[ \frac{1}{2} \widetilde{G}^{AB} \widetilde{\partial}_A \overline{\psi} \widetilde{\partial}_B \psi + \overline{\psi} v(\widetilde{q}) \psi \right], \tag{531}$$

in linear minisuperspace coordinates $\widetilde{q}^A_C \equiv (a_C, \phi_C)$ is the noncommutative action:

$$S_2^\theta[\widetilde{G}_{AB}, \psi, \overline{\psi}] = -\int_{M_\theta} d^2 \widetilde{q}_C \sqrt{-\widetilde{G}} \left[ \frac{1}{2} \widetilde{G}^{AB} \widetilde{\partial}_A \overline{\psi} \star \widetilde{\partial}_B \psi + \overline{\psi} \star v(\widetilde{q}_C) \star \psi \right],$$
$$(M_\theta \equiv (M, \star)) \tag{54}$$

and after the Lagrange variation of $\overline{\psi}$ one obtains the NWDW equation (14).

The Moyal product that occurs in action (54) has already been defined in Eq. (11) and is dropped out in the determinant of metric $\widetilde{G}_{AB}$ as well as in its contraction with other mathematical objects in action (54) due to its constant values as determined by (8). The construction of the noncommutative action (54) is well defined because the Moyal product in Eq. (54) is associative i.e. for three smooth functions $\widetilde{A}, \widetilde{B}, \widetilde{C} \in C^\infty(M_\theta)$ it satisfies the associative property [44]:

$$(\widetilde{A} \star \widetilde{B}) \star \widetilde{C} = \widetilde{A} \star (\widetilde{B} \star \widetilde{C}), \tag{55}$$

as well as for two smooth functions of compact support $\tilde{A} = \tilde{A}(\tilde{q}_C), \tilde{B} = \tilde{B}(\tilde{q}_C) \in C^\infty(M_\theta)$ it satisfies the trace property [44]:

$$\int_{M_\theta} d^2\tilde{q}_C \sqrt{-\tilde{G}}\, \tilde{A} \star \tilde{B} = \int_{M_\theta} d^2\tilde{q}_C \sqrt{-\tilde{G}}\, \tilde{B} \star \tilde{A} = \int_{M_\theta} d^2\tilde{q}_C \sqrt{-\tilde{G}}\, \tilde{A}\tilde{B}. \tag{56}$$

The noncommutative algebra (10) is in the coordinate representation therefore realized via the Moyal product (11), where the minisuperspace coordinates are the eigenvalues (Weyl symbols) of the commutative Hermitian operators of minisuperspace coordinates. This justifies the minimal replacement of the standard product in (531) with the product (11), as well as the construction of the noncommutative action (54). Minding that the noncommutative geometry, as well as the noncommutative dynamics, of the classical model of the closed universe in curvilinear minisuperspace coordinates is determined by algebra (42) we suggest that the noncommutative geometry of the corresponding quantum model may be determined by means of the diamond product, i.e. that for two minisuperspace scalar functions the following relation holds:

$$\psi_1(q^A) \Diamond \psi_2(q^B) = \psi_1(q^A) \exp\left(\frac{i\theta\varepsilon^{CD}}{2\sqrt{-G^*}} \overleftarrow{\nabla}_C \overrightarrow{\nabla}_D\right) \psi_2(q^B) =$$

$$= \psi_1(q^A)\psi_2(q^B) + \frac{i\theta\varepsilon^{CD}}{2\sqrt{-G^*}}[\partial_C \psi_1(q^A)]\partial_D \psi_2(q^B) + O(\theta^2), \left(\nabla_A \psi = \partial_A \psi \equiv \frac{\partial\psi}{\partial q^A}\right) \tag{57}$$

where the subscript "c" has been dropped out, and one has to bear in mind that the curvilinear minisuperspace coordinates $q^A$ are the eigenvalues of the commutative Hermitian coordinate operators, i.e. Weyl symbols of $\hat{q}^A$ and $\nabla_A \equiv \nabla_A(\Gamma^*)$ - is the covariant derivative. From (57) the commutation relations follow:

$$\left[q^A \stackrel{\Diamond}{,} q^B\right] \equiv \left[f_1(q^A)\Diamond f_2(q^B) - f_2(q^B)\Diamond f_1(q^A)\right]\bigg|_{f_1(q)=q^A, f_2(q)=q^B} = \frac{i\theta\varepsilon^{AB}}{\sqrt{-G^*(q)}} + O(\theta^3),$$

$$\left[\vec{e}_B \stackrel{\Diamond}{,} q^A\right] = \delta^A_B - \frac{i\theta\varepsilon^{AC}}{\sqrt{-G^*(q)}}\Gamma^{*D}_{CB}\vec{e}_D + O(\theta^2),$$

$$\left[\vec{e}_A \stackrel{\Diamond}{,} \vec{e}_B\right] = \frac{i\theta\varepsilon^{CD}}{2\sqrt{-G^*(q)}}\left[(\Gamma^{*E}_{CA}\vec{e}_E)\Gamma^{*F}_{DB}\vec{e}_F - (\Gamma^{*E}_{CB}\vec{e}_E)\Gamma^{*F}_{DA}\vec{e}_F\right] + O(\theta^3),$$

$$\left[dq^A \stackrel{\Diamond}{,} dq^B\right] = \frac{i\theta\varepsilon^{CD}}{\sqrt{-G^*(q)}} \Gamma^{*A}_{EC} \Gamma^{*B}_{FD} dq^E dq^F + O(\theta^3),$$

$$\left[q^A \stackrel{\Diamond}{,} dq^B\right] = -\frac{i\theta\varepsilon^{AD}}{\sqrt{-G^*(q)}} \Gamma^{*B}_{CD} dq^C + O(\theta^3), \tag{58}$$

where $q^A = (a,\phi)$ - are the curvilinear minisuperspace coordinates, $\vec{e}_A \equiv \frac{\partial}{\partial q^A}$ - are the tangent base vectors, i.e. orts, and $dq^A = (da, d\phi)$ - are the dual base vectors, i.e. the differentials of the curvilinear minisuperspace coordinates and $f_1(q) = q^A$, $f_2(q) = q^B$ - are the projections. From the first and last of the commutation relations (58), using:

$$\nabla_C \frac{\varepsilon^{AB}}{\sqrt{-G^*}} = -\frac{1}{2}\frac{\varepsilon^{AB} G^{*EF}(G^*_{EF,C})}{\sqrt{-G^*}} + \frac{\varepsilon^{AD}\Gamma^{*B}_{DC}}{\sqrt{-G^*}} + \frac{\varepsilon^{DB}\Gamma^{*A}_{DC}}{\sqrt{-G^*}} = 0, \left(G^*_{EF,C} \equiv \frac{\partial G^*_{EF}}{\partial q^C}\right) \tag{59}$$

there follows the relation:

$$d\left[q^A \stackrel{\Diamond}{,} q^B\right] = \frac{-i\theta\varepsilon^{AB} G^{*EF}(G^*_{EF,C})}{2\sqrt{-G^*}} dq^C + O(\theta^3) =$$

$$= -\frac{i\theta(\varepsilon^{AD}\Gamma^{*B}_{DC} - \varepsilon^{BD}\Gamma^{*A}_{DC})}{\sqrt{-G^*}} dq^C + O(\theta^3) =$$

$$= \left[q^A \stackrel{\Diamond}{,} dq^B\right] + \left[dq^A \stackrel{\Diamond}{,} q^B\right] \tag{60}$$

that may also be put in the form:

$$d\left[a \stackrel{\Diamond}{,} \phi\right] = -\frac{3i\theta}{4a^4} da + O(\theta^3) = \left[a \stackrel{\Diamond}{,} d\phi\right] - \left[\phi \stackrel{\Diamond}{,} da\right] \tag{61}.$$

It is important to note the higher powers of $\theta$ in commutation relations (58). They represent additional nontrivial noncommutative quantum corrections, so that for instance the $\theta^3$ term in the first commutation relation in (58) is noncommutative quantum correction of the first theta deformed Poisson bracket in (42), i.e.:

$$\{\!\{q^A, q^B\}\!\}_\theta = \frac{1}{i}\left[q^A \lozenge q^B\right] = \frac{\theta \varepsilon^{AB}}{\sqrt{-G^*}} -$$

$$-\frac{\theta^3 \varepsilon^{GH}\varepsilon^{EF}\varepsilon^{CD}}{24(-G^*)^{3/2}}\left(\Gamma^{*A}{}_{CE,G} - \Gamma^{*I}{}_{EG}\Gamma^{*A}{}_{CI} - \Gamma^{*I}{}_{CG}\Gamma^{*A}{}_{IE}\right)\left(\Gamma^{*B}{}_{DF,H} - \Gamma^{*J}{}_{FH}\Gamma^{*B}{}_{DJ} - \Gamma^{*J}{}_{DH}\Gamma^{*B}{}_{JF}\right) + O(\theta^5)$$

(62).

In the case of the curved 2D minisuperspace without torsion, from (57) (in which we only replace the metric $G^*_{AB}$ with the general case metric $\mathbf{G}_{AB}$), and with the identity

$$\frac{\varepsilon^{AB}}{\sqrt{-\mathbf{G}}}\frac{\varepsilon^{CD}}{\sqrt{-\mathbf{G}}} = \mathbf{G}^{AD}\mathbf{G}^{BC} - \mathbf{G}^{AC}\mathbf{G}^{BD}, \tag{621}$$

for the three smooth functions (scalar fields) $A, B, C \in C^\infty(M_\theta)$ associativity is satisfied up to the order of $\theta^3$:

$$(A \lozenge B) \lozenge C - A \lozenge (B \lozenge C) = \frac{-\theta^2}{8}\left\{\left[(\nabla_E \nabla_F A)(\nabla^E B)(\nabla^F C) - 2(\nabla^E A)(\nabla_E \nabla_F B)(\nabla^F C)\right] - \right.$$

$$\left. - \left[(\nabla_F \nabla_E A)(\nabla^E B)(\nabla^F C) - 2(\nabla^E A)(\nabla_F \nabla_E B)(\nabla^F C)\right]\right\} + O(\theta^3) =$$

$$= 0 + O(\theta^3) \tag{63}.$$

Since the minisuperspace of our model is flat, the associativity of the diamond product (57) is satisfied in all orders of $\theta$, and:

$$(A \lozenge B) \lozenge C = (\tilde{A} \star \tilde{B}) \star \tilde{C} = \tilde{A} \star (\tilde{B} \star \tilde{C}) = A \lozenge (B \lozenge C) \tag{64},$$

where $A(q) = \tilde{A}(\tilde{q}), B(q) = \tilde{B}(\tilde{q}), C(q) = \tilde{C}(\tilde{q}) \in C^\infty(M_\theta)$. Also, from (621), for the two smooth functions of compact support $A = A(q), B = B(q) \in C^\infty(M_\theta)$ the trace property is satisfied up to the order of $\theta^3$ in the case of the curved 2D minisuperspace:

$$\int_{M_\theta} d^2 q_\theta \sqrt{-\mathbf{G}}\, A \lozenge B - \int_{M_\theta} d^2 q_\theta \sqrt{-\mathbf{G}}\, B \lozenge A =$$

$$= \frac{\theta^2}{16}\int_{M_\theta} d^2 q \sqrt{-\mathbf{G}}\, R\,\mathbf{G}^{CD}\left[(\partial_C A)(\partial_D B) - (\partial_C B)(\partial_D A)\right] + O(\theta^3) =$$

$$= 0 + O(\theta^3) \tag{65},$$

where

$$dq_\theta^2 \sqrt{-\mathbf{G}} \equiv \frac{1}{2} dq^A \overset{\Diamond}{\wedge} dq^B \, \varepsilon_{AB} \sqrt{-\mathbf{G}}, \left( dq^A \overset{\Diamond}{\wedge} dq^B \equiv \frac{1}{2}\left( dq^A \Diamond dq^B - dq^B \Diamond dq^A \right) \right) \qquad (66)$$

is the measure of the minisuperspace $M_\theta = (M, \Diamond)$. It is important to see that the product (57) is not present in the metric determinant as well as in the product of this determinant with the integrand, and in the contraction of this metric with other mathematical objects in (65) and (66), what is due to the metricity condition:

$$\nabla_C \mathbf{G}_{AB} = 0 \qquad (67).$$

In our model the trace property of the diamond product (57) is satisfied in all orders of $\theta$ since:

$$\int_{M_\theta} d^2 q_\theta \sqrt{-G^*} \, A \Diamond B = \int_{M_\theta} d^2 \tilde{q} \sqrt{-\tilde{G}} \, \tilde{A} \star \tilde{B} = \int_{M_\theta} d^2 \tilde{q} \sqrt{-\tilde{G}} \, \tilde{B} \star \tilde{A} =$$
$$= \int_{M_\theta} d^2 q_\theta \sqrt{-G^*} \, B \Diamond A$$

(68),

where $\quad dq_\theta^2 \sqrt{-G^*} \equiv \frac{1}{2} dq^A \overset{\Diamond}{\wedge} dq^B \, \varepsilon_{AB} \sqrt{-G^*} = \frac{1}{2} dq^A \wedge dq^B \, \varepsilon_{AB} \sqrt{-G^*}, \, A(q) = \tilde{A}(\tilde{q})$

and $B(q) = \tilde{B}(\tilde{q})$. Starting with (64) and (68) the following noncommutative action may be constructed:

$$S_2^\theta \left[ G^*_{AB}(q), \psi, \overline{\psi} \right] = -\int_{M_\theta} d^2 q \sqrt{-G^*} \left[ \frac{1}{2} G^{*AB} \partial_A \overline{\psi} \Diamond \partial_B \psi + \overline{\psi} \Diamond v(q) \Diamond \psi \right], (M_\theta \equiv (M, \Diamond))$$

(69)

that at the finite diffeomorphism (see (6) and (7)) transforms into action (54). Action (69) can be obtained via the use of Weyl rescaling (see (49) and (52)) from the action:

$$S_2'^\Theta \left[ G'^*_{AB}(q), \psi', \overline{\psi}' \right] = -\int_{M_\Theta^\bullet} d^2 q \sqrt{-G'^*} \left[ \frac{1}{2} G'^{*AB} \overset{\bullet}{\partial}_A \overline{\psi}' \overset{\bullet}{\Diamond} \overset{\bullet}{\partial}_B \psi' + \overline{\psi}' \overset{\bullet}{\Diamond} v'(q) \overset{\bullet}{\Diamond} \psi' \right], \left( M_\Theta^\bullet \equiv \left( M, \overset{\bullet}{\Diamond} \right) \right)$$

(70)

the construction of which is enabled by the properties (64), (68), the condition of semimetricity:

$$\overset{\bullet}{\nabla}_A G'^*_{AB}(q) = 0, \qquad (71)$$

as well as the modification of the diamond product defined for instance for two minisuperspace scalar densities with:

$$\psi'_1(q^A) \overset{\bullet}{\diamond} \psi'_2(q^B) = \psi'_1(q^A)\exp\left(\frac{i\Theta^{CD}(q)}{2}\overset{\leftarrow}{\overset{\bullet}{\nabla}}_C \overset{\rightarrow}{\overset{\bullet}{\nabla}}_D\right)\psi'_2(q^B) =$$

$$= \psi'_1(q^A)\psi'_2(q^B) + \frac{i\Theta^{CD}(q)}{2}\left[\overset{\bullet}{\partial}_C \psi'_1(q^A)\right]\overset{\bullet}{\partial}_D \psi'_2(q^B) + O(\theta^2), \left(\overset{\bullet}{\nabla}_A \psi' = \overset{\bullet}{\partial}_A \psi'\right) \quad (72)$$

where $\Theta^{AB}(q) \equiv \dfrac{\Theta(q)\varepsilon^{AB}}{\sqrt{-G'^*(q)}}$ - is the cocovariantly constant 2nd rank tensorial density,

$\overset{\bullet}{\partial}_A \equiv \partial_A - \dfrac{1}{2}\rho w'_A,\ w'_A = -2\Omega^{-1}\partial_A\Omega;\ (\Omega^2(q) = a),\ G'^*_{AB}(q) = \Omega^{-2}(q)G^*_{AB}(q)$ and

$\Theta(q) = \Omega^{-2}(q)\theta$ -is the scalar density with Weyl weight $\rho = 2$. The noncommutative action (70) completely determines the noncommutative dynamics of the quantum minisuperspace model of the closed universe.

Let us consider now the effect of the noncommutativity of the minisuperspace on the classical cosmology of the closed universe described by the FRW metrics determined by the space-time interval:

$$ds^2 = \sigma^2\left[-N'^2(t)dt^2 + a^2(t)d\Omega_3^2\right], \quad (73)$$

which results from the application of the Weyl rescaling [44]:

$$N'(t) = \Omega^{-2}(q)N(t), \left(\Omega^2(q) = a(t)\right) \quad (74)$$

to (1). The choice of the Weyl rescaling factor $\Omega(q)$ in (74) appears adequate since it enables the flat minisuperspace to remain conformally flat as well [54]. We demonstrated earlier that the noncommutativity of the minisuperspace under the H-H condition results in the most probable creation of the closed universe with the values of the real parameters of $\alpha > 0$ and $\beta = 0$ under the gauge condition $N^+_{\theta a}$ (see (307)), that determine the stable matter potential with a minimum in $\phi = 0$ (see (2)). Now we shall consider the classical cosmology of this most probable universe. If we apply the Weyl rescaling (see (49) and (52)) to the noncommutative hamiltonian (38) with the above mentioned values of the matter field parameters, we obtain the following noncommutative hamiltonian:

$$H'^\Theta \equiv N'H'_\Theta = \frac{N'}{2}\left[-\frac{p_a^2}{a} + \frac{p_\phi^2}{a^3} + \alpha a^3 \text{ch}2\phi - a - \frac{\Theta\alpha}{4}\left(-p_a a\,\text{sh}2\phi + p_\phi \text{ch}2\phi\right)\right] \approx 0$$

(75),

where the subscript "c", denoting that the minisuperspace coordinates and momenta satisfy the Poisson algebra (37), has been dropped out for simplicity. Applying the Legandre transformations to (75) we obtain the noncommutative Lagrangian:

$$L'_\Theta\left[a, \frac{da}{dt}, N'\right] = \frac{N'}{2}\left[-\frac{a}{N'^2}\left(\frac{da}{dt}\right)^2 - \left(\alpha - \frac{\Theta^2 \alpha^2}{64}\right)a^3 + a\right]. \tag{76}$$

Solving the equations of motion obtained from (76), taking care of the gauge condition $N'=1$ and the initial conditions $\left.\frac{da}{dt}\right|_{t=0} = 0, \left.\frac{d^2 a}{dt^2}\right|_{t=0} = \alpha\, a(0)$ we obtain the Lorentz 4-metric determined by the space-time interval:

$$ds^2 = -dt^2 + \left(\frac{1}{\alpha} + \frac{\theta^2 \alpha}{64}\right)\mathrm{ch}^2\left(\sqrt{\alpha}\, t\right)d\Omega_3^2, (\sigma = 1) \tag{77}$$

For $\theta = 0$ this reduces to the metrics of the de Sitter space-time, the symmetry of which is described by the de Sitter group SO(1,4) [38], [39]. The scalar curvature of the thus $\theta$-deformed de Sitter space-time is:

$$^4R(t) = 4\Lambda\left[1 - \frac{\theta^2 \Lambda^2}{1152\left(1 + \frac{\theta^2 \Lambda^2}{576}\right)\mathrm{ch}^2\left(\sqrt{\Lambda/3}\, t\right)}\right], \tag{78}$$

while the scalar curvature of the 3D subspace is:

$$^3R(t) = \frac{2\Lambda}{\left(1 + \frac{\theta^2 \Lambda^2}{576}\right)\mathrm{ch}^2\left(\sqrt{\Lambda/3}\, t\right)}, \tag{79}$$

where $\Lambda = 3\alpha$. From (77-79), the square of the Hubble parameter equals:

$$H^2(t) = \frac{\Lambda_{eff}(t)}{3} - \frac{1}{a^2(t)}, \tag{80}$$

where $\Lambda_{eff}$ is the effective cosmological "constant", determined by the expression:

$$\Lambda_{\text{eff}}(t) = \Lambda\left[1 - \frac{\theta^2 \Lambda^2}{\left(576 + \theta^2 \Lambda^2\right)\text{ch}^2\left(\sqrt{\Lambda/3}\,t\right)}\right]. \tag{81}$$

At cosmic time $t = 0$, the Hubble parameter equals zero, i.e.

$$H^2(0) = \frac{1}{a^2(t)}\left[\frac{da(t)}{dt}\right]^2\bigg|_{t=0} = 0, \tag{82}$$

and the value of the cosmological "constant" is:

$$\Lambda_{\text{eff}}(0) = \frac{\Lambda}{1 + \frac{\theta^2 \Lambda^2}{576}}. \tag{83}$$

For $t \to \infty$ from (78-81) there follows that:

$$^4R \to 4\Lambda,\ ^3R \to 0,\ H^2 \to \Lambda/3,\ \Lambda_{\text{eff}} \to \Lambda, \tag{84}$$

This means that the effects of noncommutativity get weaker with time, and become negligible at great separations. In other words, the $\theta$-deformed closed universe accelerates its expansion, to asymptotically reach the geometry of the de Sitter space-time with the flat 3D subspace in infinite future. Also, assuming the validity of the gauge condition $N'=1$ and initial conditions $a(0)=0$, $\frac{da}{d\tau}\bigg|_{\tau=0} = \sqrt{1 + \frac{\theta^2 \alpha^2}{64}}$, after applying the Wick rotation to (76), and obtaining the corresponding equations of motion from this non-commutative Lagrangian, we obtain the Euclidian 4-metric determined by the space-time interval:

$$ds_E^2 = d\tau^2 + \left(\frac{1}{\alpha} + \frac{\theta^2 \alpha}{64}\right)\sin^2\left(\sqrt{\alpha}\,\tau\right)d\Omega_3^2, \quad (\sigma = 1). \tag{85}$$

For $\theta = 0$ this reduces to the metric of the maximum symmetric 4-sphere (Hartle-Hawking gravitational instanton) of the radius $1/\sqrt{\alpha}$, whose symmetry is described by the SO(5) group [38], [39]. From (85), the semiclassical noncommutative Hartle-Hawking (H-H) wave functions that corresponds to this 4-geometry are of the form:

$$\psi_\theta'^{\pm}(a'' = a(1)) \approx \exp\left\{\pm\frac{1}{3\alpha}\left(1 + \frac{\theta^2 \alpha^2}{64}\right)^{3/2}\left[1 - \alpha a''^2\left(1 + \frac{\theta^2 \alpha^2}{64}\right)^{-1}\right]^{3/2}\right\} \tag{86}.$$

From Eq. (86) we see that the noncommutativity parameter $\theta$ increases (for the "+" sign) or decreases (for the "–" sign) the two corresponding standard semiclassical H-H tunnelling amplitudes [38], [39]. So, from this consideration we may now conclude that the canonical noncommutativity prefers the creation of the theta deformed de Sitter universe rather by $\psi_\theta'^+$ than by $\psi_\theta'^-$. For $\theta = 0$, $\psi_\theta'^+$ corresponds to the geometry of filling in the three-sphere with more than half of a four-sphere of the radius $1/\sqrt{\alpha}$. This result is not unexpected because the noncommutative parameter $\theta$ bounds the cosmic scale factor from below.

**Conclusion**

A number of results follow from our model of the universe. The main result is probably that the canonical noncommutativity of the minisuperspace chooses the creation from nothing to the classical universe with stable matter potentials under the gauge condition $N_{\theta a}^+$ (see (307)), the universe with the most stable potential being most probable to create. The classical universes with unstable matter potentials under the same gauge conditions (307) do not create from nothing, because their corresponding wave functions (3013) do not decohere in the cosmic large-scale structure. This, together with the appropriate restrictions on the free parameters of the model and the Hartle-Hawking condition, chooses only the most probable semiclassical wave function of the universe from the class of semiclassical particular solutions of the noncommutative Wheeler-DeWitt equation. This is why in the multiverse picture the canonical noncommutativity of the minisuperspace appears as the natural superselection rule, making the need for the anthropic principle superfluous. Also, we conjectured that the noncommutativity of curvilinear minisuperspace coordinates (the cosmic scaling factor and the matter field) is realized by the modification of the Moyal product. We show that such construction of noncommutativity at the classical level leads to smaller values of the cosmological "constant" in the early phases of the theta deformed de Sitter universe created with greatest probability, whose further evolution with the passing of cosmic time determines the increase of this "constant". With the flow of cosmic time the effects of noncommutativity get weaker, the expansion of the universe accelerates and the universe asymptotically reaches the standard de Sitter space-time geometry with 3D flat subspace in infinite future. The significance of this kind of noncommutativity is thus not limited to early phases of the evolution of the universe where it prefers creation by semiclassical tunneling amplitude $\psi_\theta'^+$ (see (86)), but also influences the long-scale structure and future evolution of the observable universe due to the dependence of its 4-geometry on cosmic time.